\documentclass[submission,copyright,creativecommons]{eptcs}
\providecommand{\event}{FROM 2024} 

\usepackage{iftex}

\ifpdf
  \usepackage{underscore}         
  \usepackage[T1]{fontenc}        
\else
  \usepackage{breakurl}           
\fi

\usepackage{xcolor}
\usepackage[utf8]{inputenc}     
\usepackage{graphicx}           
\usepackage{indentfirst}        
\usepackage[nottoc]{tocbibind}  
\usepackage{hyperref}
\usepackage{cite}
\usepackage{algorithm}
\usepackage{algpseudocode}
\usepackage{listings}
\usepackage{amsmath}
\usepackage{array}
\usepackage{amssymb}
\usepackage{multicol}
\usepackage[normalem]{ulem}
\usepackage{xspace}

\newcommand{\mynull}{\texttt{null}\xspace}
\newcommand{\allocate}{\texttt{allocate}\xspace}
\newcommand{\deallocate}{\texttt{deallocate}\xspace}
\newcommand{\create}{\texttt{create}\xspace}
\newcommand{\destroy}{\texttt{destroy}\xspace}
\newcommand{\Owner}{\texttt{Owner}\xspace}
\newcommand{\Pointer}{\texttt{Pointer}\xspace}
\newcommand{\Value}{\texttt{Value}\xspace}

\newcommand{\pre}{\texttt{pre}\xspace}
\newcommand{\post}{\texttt{post}\xspace}
\newcommand{\myglobal}{\texttt{global}\xspace}
\newcommand{\invalid}{\texttt{invalid}\xspace}

\title{Static Analysis Framework for Detecting Use-After-Free Bugs in C++}
\author{Vlad-Alexandru Teodorescu
\institute{Faculty of Computer Science\\
Alexandru Ioan Cuza University of Iasi, Romania}
\email{teodorescu.vlad@yahoo.com}
\and
Dorel Lucanu
\institute{Faculty of Computer Science\\
Alexandru Ioan Cuza University of Iasi, Romania}
\email{dorel.lucanu@gmail.com}
}
\def\titlerunning{Static Analysis Framework for Detecting Use-After-Free Bugs in C++}
\def\authorrunning{Vlad-Alexandru Teodorescu \& Dorel Lucanu}
\begin{document}
\maketitle

\begin{abstract}
Pointers are a powerful, but dangerous feature provided by the C and C++ programming languages, and incorrect use of pointers is a common source of bugs and security vulnerabilities. Making secure software is crucial, as vulnerabilities exploited by malicious actors not only lead to monetary losses, but possibly loss of human lives. Fixing these vulnerabilities is costly if they are found at the end of development, and the cost will be even higher if found after deployment. That is why it is desirable to find the bugs as early in the development process as possible. We propose a framework that can statically find use-after-free bugs at compile-time and report the errors to the users. It works by tracking the lifetime of objects and memory locations pointers might point to and, using this information, a possibly invalid dereferencing of a pointer can be detected. The framework was tested on over 100 handwritten small tests, as well as 5 real-world projects, and has shown good results detecting errors, while at the same time highlighting some scenarios where false positive reports may occur. Based on the results, it was concluded that our framework achieved its goals, as it is able to detect multiple patterns of use-after-free bugs, and correctly report the errors to the programmer.
\end{abstract}

\section{Introduction}

Pointers are a powerful, but dangerous feature provided by the C and C++ programming languages, and incorrect use of pointers is a common source of bugs and security vulnerabilities. Most new languages lack pointers or severely restrict their capabilities, thus eliminating these problems and providing memory safety. Nonetheless, many C \& C++ programs work with pointers safely and they are still considered a very useful feature of the language. Programmers who safely work with pointers maintain an internal model of when memory accessed through those pointers should be allocated and subsequently freed. Commonly applied models include garbage collection, Resource Acquisition Is Initialization (RAII), and smart pointers. 

However, because the chosen model is frequently not documented in the program, it might not be well understood when modifying the source code, and errors can appear. These can lead to various problems such as program crashes, unpredictable behavior, or security vulnerabilities. Some infamous examples of critical vulnerabilities caused by memory bugs are:
\begin{itemize}
    \item Heartbleed (CVE-2014-0160) \cite{heartbleed} - exposed secrets in the popular OpenSSL library. It affected over 66\% of the active sites on the Internet.
    \item BlueKeep (CVE-2019-0708) \cite{bluekeep} - Remote Code Execution in Microsoft's RDP. It affected all unpatched Windows versions.
    \item EternalBlue (CVE-2017-0144) \cite{eternalblue} - Remove Code Execution in Windows Microsoft's SMB. It affected all unpatched Windows versions. It was used to carry out cyber attacks which caused major damage, like WannaCry \cite{wannacry} or NotPetya \cite{notpetya}
\end{itemize}

Making secure software is crucial, as vulnerabilities exploited by malicious actors not only lead to monetary losses, but possibly loss of human lives \cite{catastrofa}. Fixing these vulnerabilities is costly if they are found at the end of development, and the cost will be even higher if found after deployment. That is why it is desirable to find the bugs as early in the development process as possible. One class of tools that can help programmers identify software defects early is static analysis tools. These tools analyze source code without executing it, and can even point out bugs while the programmer is still writing the code.

One common bug in C++ that can cause vulnerabilities is the use-after-free pattern. They occur when a program continues to use a pointer after the memory it points to has been freed. This usually causes program crashes, but in the worst cases, it may lead to critical vulnerabilities such as those mentioned earlier.

Identifying this type of bug is not trivial, as C++ is a complex language that gives the programmer a lot of control over memory, without many constraints. Access to on-demand allocation and deallocation of memory, combined with pointer arithmetic and the existence of functions that can affect memory outside of their scope make tracking pointer operations at compile time difficult, and sometimes even impossible. In addition to the difficulty of identification by static analyzers, this type of bugs are also very hard to identify during code reviews, even by experienced programmers.

\paragraph{Contribution} This paper proposes a framework intended to provide sound static analysis by using semantic information that can be inferred from the source code and some information explicitly provided by the programmer. In particular, the framework aims to track the lifetime of all memory regions referenced by pointers during the program's execution and ensure that dereferenced pointers' pointed-to memory is valid. 

Firstly, a simplified model is extracted from the target program to facilitate the analysis. Types and instructions are separated into multiple categories depending on their properties and possible effects on memory. Then, the identification of errors is done through dataflow analysis on the control flow graph of the simplified program. The use of tools from the Clang ecosystem facilitates all these steps.

\paragraph{Paper Structure} This paper is split into 4 sections, each describing theoretical aspects or implementation details of the framework. Section 2 describes other approaches for detecting memory bugs. Section 3 presents all the steps of the analysis process used for detecting and reporting the errors. Section 4 describes some of the implementation details of the steps presented in the previous section and the results of several experiments conducted to evaluate the framework's performance.

\section{Related Work}

Since software security is crucial, there have been many projects that try to detect and prevent vulnerabilities before they are available to the public and can be exploited. The classes of bugs detected and the concepts used to detect them differ from project to project, but most of them fall under 3 main categories: static analyzers, dynamic analyzers, and hybrid approaches

\subsection{Static Analyzers}

Static analyzers are tools that check different properties of programs by only looking at their source code, without the need to execute them.

Herb Sutter describes in \cite{lifetime} an approach to identify use-after-free bugs in C++ through static analysis and is the main inspiration for this framework and paper. 

Multiple other tools aim to find different types of memory bugs through static analysis. The Clang static analyzer supplements the built-in warnings of the compiler with the same name with over 100 powerful checkers \cite{clangcheck} that try to detect both general C++ bugs, as well as application-specific ones. The other 2 main C++ compilers, GNU GCC and Microsoft Visual C++ have similar static analysis tools.

In addition to the warnings produced by the compilers, IDEs have their own tools looking for errors. Jetbrains CLion provides tens of different warnings \cite{clion} and also integrates with other static analysis tools in order to have as much coverage as possible when looking for bugs. Other popular IDEs such as Microsoft Visual Studio or XCode have similar approaches.

\subsection{Dynamic Analyzers}

Dynamic analysis is the process of evaluating a program by executing it and observing its behavior. In the case of C++ this usually involves tracking memory management and access, as well as concurrency, to ensure no errors are present.

Valgrind \cite{valgrind} is a powerful tool suite that is primarily used for detecting memory management and threading bugs in C++ applications. It includes various tools that can identify memory leaks, invalid memory access or mismanagement of memory. Valgrind emulates the execution of programs, so it has complete control over the low-level operations. Because of this, it is able to provide detailed information about potential errors in the program, as well as their causes.

Address Sanitizer \cite{asan} is another popular dynamic analyzer that can detect memory errors. It employs a specialized memory allocator and code instrumentation to accurately detect bugs at their point of occurrence. It is currently the most widely used tool due to it having the smallest performance impact, while still being accurate. It is also integrated in all major C++ compilers.

\subsection{Hybrid Approaches}

There are a few tools that take a hybrid approach by combining both static and dynamic analysis, and one of the most popular and efficient uses of this is fuzz testing \cite{fuzz}. A fuzzer uses both static analysis to look at the source code, and dynamic analysis to observe the flow of execution in order to generate sets of inputs that cover as many branches as possible. One of the more popular fuzzers is AFL++ \cite{afl}. Then, these inputs are fed to the program, and dynamic analyzers are used to find possible errors during the execution. The wide coverage of inputs generated by the fuzzer helps find errors that happen very rarely, which may not be caught when using a hand-written set of tests.

\section{The Proposed Framework}

The analysis process consists of several steps that will be described in this section. Firstly, the language used to describe the model extracted from the input C++ program is defined. Next, the operations that transform a C++ program into the form used by the framework are presented. Finally, we describe the analysis process, what information it uses, and how it is obtained.

\subsection{The Language}

The first step to facilitate the static analysis is to extract from a C++ program a model expressing how the program interacts with memory. The language we will use is based on C++ \cite{cppstd}, with some modifications to the operations that manipulate memory. The additional syntax needed to manipulate the object's lifetime is described in Figure~\ref{fig:syntax}.

\begin{figure}[htb]
    \centering
\[
\begin{array}{rll>{\raggedright\arraybackslash}p{5cm}}
\langle\text{instructions}\rangle ::= & \ldots & &  \\
| & \langle\text{var}\rangle = \allocate(\langle\text{exp}\rangle); & & \text{ allocation} \\
| & \langle\text{var}\rangle = [\langle\text{exp}\rangle]; & & \text{ lookup} \\
| & [\langle\text{exp}\rangle] = \langle\text{exp}\rangle; & & \text{ mutation} \\
| & \deallocate(\langle\text{exp}\rangle); & & \text{ deallocation} \\
| & \create(\langle\text{var}\rangle, \langle\text{type}\rangle); & & \text{variable creation} \\
| & \destroy(\langle\text{var}\rangle); & & \text{variable destruction} \\
\langle\text{type}\rangle ::= & \Owner & & \\
| & \Pointer & & \\
| & \Value & & \\
\end{array}
\]
\vspace{-3ex}
    \caption{Additional syntax for handling the lifetime of objects}
    \label{fig:syntax}
\vspace{-2ex}
\end{figure}

The computational state contains two components: a store, which maps variables into addresses, and the memory, which maps addresses into values.
\begin{align*}
\text{Values} &= \text{Integers} \cup \text{Atoms} \cup \text{Addresses} \\
\text{Memory} &= \bigcup_{\text{A} \subseteq \text{Addresses}} (\text{A} \to \text{Values})\displaybreak[0]\\
\mynull & \in \text{Atoms}\displaybreak[0]\\
\text{Stores}_V &= \text{V} \to \text {Addresses}\\
\text{States}_V &= \text{Stores}_V \times \text{Memory}
\end{align*} 
where $V$ is a finite set of variables.

The behavior of the new instructions is defined by a transition relation $\rightsquigarrow$ between configurations, which can be:
\begin{itemize}
    \item \textbf{nonterminal:} an instruction-state pair $\langle i, (s,m) \rangle$, where $(s,m)$ is a State, $FV(i) \subseteq dom(s)$ ($FV(i)$ is the set of free variables in instruction $i$);
    \item \textbf{terminal:} a state $(s,m)$ or \textbf{error}.
\end{itemize}
The semantics of the new instructions is defined below. Here $[f|x:y]$ denotes the function that maps $x$ into $y$ and all other $x' \in dom(f)$ into $f(x')$. The notation $f|_S$ means the restriction of the function $f$ to the domain $S$. $dom(f)$ is the domain of the function $f$ and $val(e)$ is the value of expression $e$.

\begin{itemize}
    \item Allocation:\\
    $\langle v = \allocate(e), (s,m) \rangle \rightsquigarrow  
    (s, [m | s(v):a | a:null | a+1:null | \ldots | a+val(e)-1:null])$\\
    where $a, a+1, \ldots , a+val(e)-1 \in \text{Addresses} - dom(m)$

    \item Lookup:
    \\
    $ \begin{aligned}
        \langle v = [e], (s,m) \rangle &\rightsquigarrow (s, [m | s(v):m(val(e))]) &&\textrm{if~~} val(e) \in dom(m)\\
    \langle v = [e], (s,m) \rangle &\rightsquigarrow \textbf{error} && \textrm{if~~} val(e) \notin dom(m)
    \end{aligned}$
    \item Mutation: 
    \\
    $ \begin{aligned}
    \langle [e] = e', (s,m) \rangle & \rightsquigarrow (s, [m | val(e):val(e')]) &&
    \textrm{if~~} val(e) \in dom(m)\\
    \langle [e] = e', (s,m) \rangle &\rightsquigarrow \textbf{error} && \textrm{if~~} val(e) \notin dom(m)
    \end{aligned}$

    \item Deallocation:
    \\
    $\langle \deallocate(e), (s,m) \rangle \rightsquigarrow (s, m|_{dom(m) - A})$
    \\
    \phantom{deallocate} if $val(e)$ is the first address in a set of addresses returned by an \allocate instruction before the call to \deallocate and $val(e) \in dom(m)$
    \\
    $ \begin{aligned}
    \langle \deallocate(e), (s,m) \rangle & \rightsquigarrow \textbf{error}  && \textrm{otherwise}
    \end{aligned}$
    \\[1ex]
    where $A$ is the set of Addresses returned by the \allocate instruction.

    \item Variable Creation:
    \\
    $\langle \create(v,t), (s,m) \rangle \rightsquigarrow ([s|v:a],[m|a:null]) $
    \\
    where $a \in \text{Addresses} - dom(m)$;

    \item Variable Destruction:
    \\
    $\langle \destroy(v), (s,m) \rangle \rightsquigarrow (s|_{dom(s)-v},m|_{dom(m)-s(v)}$
    
\end{itemize}

The allocation instruction activates and initializes the required cells in the heap. The only requirement for these cells is that they were previously inactive and are consecutive. The starting address is unspecified.

The Lookup, Mutation, and Deallocation operations cause memory errors (indicated by the terminal configuration \textbf{error}) whenever an invalid address is dereferenced or deallocated. This would correspond to a crash when running a C++ program.

The \create and \destroy instructions are used to emulate the behavior of stack variable allocation and deallocation when entering or exiting a scope in C++.
\vspace*{-2ex}

\paragraph*{Types}

There are 3 classes of types based on their properties: \Owner, \Pointer, \Value.

An \Owner is a variable that owns a zone of memory. This means it can use all four memory management functions from our language on the memory it owns. During the variable's creation, it allocates some memory and during its destruction it deallocates it. Some functions may alter the \Owner, thus changing the memory zone managed by the variable to another one. Dereferencing an \Owner is always valid. (for example, a type classified as Owner is \texttt{std::vector})

A \Pointer is a variable that points to some memory it does not own. It does not allocate or deallocate any memory. This means it can only use the Lookup and Mutate instructions on the memory it points to. All use-after-free errors happen when dereferencing this class of variables, as they can still point to a memory zone that has already been deallocated by an \Owner. A \Pointer that produces an error when dereferenced is called invalid.

A variable is classified as a \Value if it is neither an \Owner, nor a \Pointer. All variables that do not interact with heap memory fall into this category.
\vspace*{-2ex}

\paragraph*{Annotations}

In addition to the modifications to instructions, we also extend the function definition syntax to be able to make annotations that represent preconditions about the lifetime of the input or output parameters of the function and postconditions about the lifetime of its outputs.

We define the lifetime of a memory zone as the sequence of instructions in the program during which it is valid to dereference it. Formally, the address $a$ is alive if none of its lookup, mutation, or deallocation transition to the \textbf{error} configuration.

The lifetime of a memory zone begins when the \allocate instruction activates the corresponding memory addresses and ends when the \deallocate instruction deactivates them.

The syntax of the annotations is:
\[
\begin{array}{rll>{\raggedright\arraybackslash}p{5cm}}
\langle\text{annotated\_func}\rangle ::= & \langle \text{annotation} \rangle \langle \text{func\_definition} \rangle & & \\
\langle\text{annotation}\rangle ::= & \pre((\langle \text{var} \rangle, \{ \langle \text{lifetime} \rangle * \})+) & & \text{precondition} \\
| & \post((\langle \text{var} \rangle, \{ \langle \text{lifetime} \rangle * \})+) & & \text{postcondition}\\
\langle\text{lifetime}\rangle ::= & \langle \text{var} \rangle & &  \\
| & \mynull & & \\
| & \myglobal & & \\
| & \invalid & & \\
\end{array}
\]

Let $p$ be a function parameter with the precondition $\pre(p, S)$ attached. This means that $p$ must have the same lifetime as at least one element in $S$ when the function is called. For example, when calling the function defined in listing \ref{lst:expre}, the lifetime of the parameter $z$ has to be equal to either that of $x$ or $y$, otherwise, an error is raised.

To attach a postcondition to the return value of a function, we will use the function's name. Let $f$ be a function with postcondition $\post(f, S)$. This means that the return value of $f$ must have the same lifetime as at least one element in $S$. For example, the value returned by the function defined in listing \ref{lst:expost} has to have a lifetime equal to that of $x$, or be null.

\noindent\begin{minipage}{.48\textwidth}
\begin{lstlisting}[caption=A function with a precondition,frame=tlrb,language=C++,label={lst:expre}]
pre(z,{x,y})
Pointer f(Pointer x,Pointer y,
Pointer z)
{
...
}
\end{lstlisting}
\end{minipage}\hfill
\begin{minipage}{.48\textwidth}
\begin{lstlisting}[caption=A function with a postcondition,frame=tlrb,language=C++,label={lst:expost}]
post(g, {x,null})
Pointer g(Pointer x)
{
...
}
\end{lstlisting}
\end{minipage}

\subsection{Transforming a C++ Program}

We define a set of rules that will help the framework transform the input C++ program into our proposed simplified language. 

\subsubsection{Variable Types} \label{sec:types}

A variable is classified as an \Owner if its type satisfies any of the following conditions:
\begin{itemize}
    \item it satisfies the standard Container requirements and has a user-provided destructor - for example, \texttt{std::vector};
    \item it provides the dereference operator and has a user-provided destructor - for example, \texttt{std::unique\_ptr};
    \item it has a data member or base class of type \Owner.
\end{itemize}

A variable is classified as a \Pointer if its type is not an \Owner and satisfies any of the following conditions:
\begin{itemize}
    \item it satisfies the standard Iterator requirements;
    \item Is trivially copyable, is copy constructible and assignable, and provides the dereference operator.
    \item it has a data member or base class of type \Pointer;
    \item it is a capture by reference inside a lambda function.
\end{itemize}

A variable is classified as a \Value if it is neither an \Owner nor a \Pointer. All variables that do not interact with heap memory fall into this category.

\subsubsection{Instructions}

Control-flow instructions, assignments, expressions, and most other instructions remain unchanged. The only cases where explicit transformations are needed are memory-related instructions.

Raw pointer arithmetic makes static tracking of memory zones variables are pointing to very difficult, sometimes even impossible. That is why our framework forbids it, as the analysis may give imprecise, or even wrong results because of these instructions.

Manual memory allocation through \texttt{new} is considered a special case of creating a global \Owner variable. Manual memory deallocation through \texttt{delete} corresponds to the destruction of this variable.

\[
\begin{array}{rl}
\texttt{p = new int;} & \longrightarrow \quad \begin{array}{l}
\text{\create(global\_owner, \Owner);} \\
\text{global\_owner = \allocate(4);} \\
\text{p = global\_owner;}
\end{array}
\end{array}
\]

\[
\begin{array}{rl}
\texttt{delete }\text{p;} & \longrightarrow \quad \begin{array}{l}
\text{\destroy(global\_owner);} \\
\text{\deallocate(global\_owner);}
\end{array}
\end{array}
\]
In C++, when a scope begins, all local variables are implicitly allocated on the stack. In a similar manner, when a scope ends, all local variables are implicitly deallocated. In our programming language, this is done explicitly.

\[
\begin{array}{rl}
\begin{array}{l}
    \text{\{} \\
    \texttt{~~int }\text{x;} \\
    \texttt{{}~~\ldots} \\
    \text{\}}
\end{array}
& 
\longrightarrow \quad 
\begin{array}{l}
    \text{\{} \\
    \text{~~\create(x, \Value);} \\
    \text{~~\ldots} \\
    \text{~~\destroy(x);} \\
    \text{\}}
\end{array}
\end{array}
\]

\subsubsection{Functions} \label{sec:funcCond}

Functions can be explicitly annotated by the programmer with preconditions and postconditions about the lifetime of their \Pointer parameters. 
If these are not explicitly mentioned, then the following default conditions will be enforced:
\begin{description}
    \item[Precondition] No \Pointer arguments can point to a non-const global \Owner or a local \Owner being passed by a non-const reference to the same function call. Also, no two \Pointer arguments should point to the same non-const \Owner.
    \item[Postcondition] If the function returns a \Pointer variable, its lifetime should be at least as long as one of the arguments. 
\end{description}

\subsubsection*{Example program}

Following the rules above, an example transformation of a section of a program is presented in listings \ref{lst:cppprog} and \ref{lst:transprog}.

\noindent\begin{minipage}{.48\textwidth}
\begin{lstlisting}[caption=Example C++ program,frame=tlrb,language=C++,label={lst:cppprog}]
int x;
int* p;
...
if (x == 2) p = &x;
else {
    int y;
    p = &y;
}
*p = 3;
\end{lstlisting}
\end{minipage}\hfill
\begin{minipage}{.48\textwidth}
\begin{lstlisting}[caption=Transformed C++ program,frame=tlrb,language=C++,label={lst:transprog}]
create(x,Value);
create(p,Pointer);
...
if (x == 2) p = &x;
else {
    create(y,Value);
    p = &y;
    destroy(y);
}
*p = 3;
\end{lstlisting}
\end{minipage}

\subsection{Analysis} \label{sec:rules}

After the steps above, the framework will perform static intraprocedural analysis, on each function definition, to enforce the following rules:

\begin{enumerate}
    \item It is an error to dereference an invalid \Pointer.
    \item It is an error to dereference an \Owner that was moved from (transferred ownership to another variable).
    \item It is an error to use raw pointer arithmetic.
    \item It is an error to pass a \Pointer as a function argument if it is invalid, or violates the preconditions of the function.
    \item It is an error to return a \Pointer from a function that is invalid or violates the postconditions of the function.
\end{enumerate}

To enforce rule 3., we will always report a possible error when pointer arithmetic is present in the analyzed program. Let $p$ be a pointer and $v$ another variable. The following expressions will produce an error:
\begin{multicols}{6}
\begin{itemize}
    \item p+v
    \item v+p
    \item p++
    \item ++p
    \item p-v
    \item v-p
    \item p- -
    \item - -p
    \item p+=v
    \item p-=v
    \item p[x]
    \item x[p]
\end{itemize}
\end{multicols}

While rule 3. can be easily enforced through the analysis of the AST of the program and the types of variables, the rest of the rules need more complex, path-sensitive analysis. They all result in the same behavior of the program: reaching the \textbf{error} configuration after a Lookup of Mutate instruction.

A way to enforce these rules is by maintaining all possible locations a pointer might point to, and whether they are valid or not. We will use the notions of \textit{points-to-map} (pmap) and \textit{points-to-set} (pset) to maintain this information.

The \textit{points-to-set} of a variable $v$  at instruction $i$ in the program is the set of all possible memory zones it may point to at this moment during the execution of the program. This set can contain any of these elements:
\begin{description}
    \item[var] - this means that $v$ currently points to the address of local variable;
    \item[global] - $v$ currently refers to a static variable or a memory owned by a const static \Owner;
    \item[o'] - $v$ points to the address of an object owned by \Owner $o$;
    \item[o'' (etc.)] - $v$ refers to an object owned by an object owned by $o$. This is used in the case of hierarchies of \Owner objects;
    \item[null] - $v$ currently is null;
    \item[invalid] - $v$ points to invalid memory.
\end{description}

The \textit{points-to-map} at instruction $i$ is a mapping between local variables that exist at the current moment in execution and their psets.

A variable $v$ is considered invalid at instruction $i$ if $\invalid \in pmap_i(v)$. (see for example $p$ on the last line in listings \ref{lst:cppprog} and \ref{lst:transprog})

We have chosen to calculate $pmap_i$ for all instructions using Dataflow Analysis, as some form of path-sensitive analysis is required to correctly determine the points-to-sets.

\subsection{Instantiation of the Dataflow Analysis Framework}

Let $\mathcal{F} = (\mathcal{V}, \mathcal{E}, e)$ be the control flow graph of the program we want to analyze. To calculate all the points-to-maps at each node in the graph, we will solve the following dataflow system~\cite{progAnalysis}:
\begin{itemize}
    \item $L = \mathcal{V}$. The program labels are the labels of the nodes.
    \item $E = e$. We will start the analysis from the initial nodes of the CFG.
    \item $F = \mathcal{E}$. The flow of the analysis is determined by the edges in the graph.
    \item $D = \text{Variables}\rightharpoonup 2^{\text{PsetEntries}}$ is the analysis domain, 
    where $\text{PsetEntries} = \text{Variables} \cup \text{Variables'} \cup \{ null, invalid, global\}$ 
    ,
    and $\text{Variables'} = \{v', v'',\ldots|v \in \text{Variables}\}$.
\\
    The information we maintain is the points-to-map at each node.
    
    \item $\sqsubseteq $ is the partial order defined as follows: $f \sqsubseteq g$ iff $dom(f) \subseteq dom(g)$ and for all $v \in dom(f)$, $f(v) \subseteq g(v)$.
    \item  ${\bot} = \emptyset$ is the smallest element, and the initial value associated with $e$ is $\iota = \emptyset$.
\end{itemize}

The transfer function is the most important part of this system and affects the $pmap$ in different ways depending on the instruction in the node. 

$\varphi_i$ is the transfer function corresponding to instruction $i$ and takes as input $d$ which is the union of all $pmaps$ of predecessor nodes in the CFG and produces a new $pmap$ after the effects of instruction $i$.

At different moments during the analysis, $pmap$ entries will be invalidated. We define the invalidation of a set of variables $S$ as\\[1ex]
\centerline{$ 
\text{invalid}_S(d) = [d|x:(d(x) - \{v,v',\ldots \}) \cup \{ invalid \}]
$}\\[1ex]
where $x$ such that $d(x) \cap \{v,v',\ldots \} \not= \emptyset$ for all $v \in S$.

\begin{description}
    \item[Variable creation] - when creating a variable, a new entry is added to the $pmap$. Let $i$ be the instruction $\create(v,t)$. Then,\\[1ex]
\centerline{$
        \varphi_i(d) = 
        \begin{cases}
        [d|v:\{v'\}] & \text{if} \; t = \Owner \\
        [d|v:\{invalid\}] & \text{if} \; t = \Pointer \\
        d & \text{if} \; t = \Value
        \end{cases}
$}  
    \item[Variable destruction] - when destroying a variable, its entry is removed, and all other $pset$s that may refer to it are invalidated. Let $i$ be the instruction  $\destroy(v)$. Then,\\[1ex]
\centerline{$ \varphi_i(d) = \text{invalid}_v(d|_{dom(d) -\{v\}})$}

    \item[Mutation of owner] - when a non-const use of a local \Owner occurs, we will invalidate all pointers to it, making the conservative assumption that any use might cause reallocation. Let $i$ be $f(o)$. Then,\\[1ex]
\centerline{$ \varphi_i(d) = \text{invalid}_o(d)$}

    \item[Copying] - when a copy happens, all the entries in the destination's $pset$ will be replaced by all the entries in the source's set. Let $i$ be the instruction be $v = u$. Then,\\[1ex]
\centerline{$ \varphi_i(d) = [d|v:d(u)] $}

    \item[Moving] - when a move happens, all the entries in the destination's $pset$ will be replaced by all the entries in the source's set. In addition to this, the source's $pset$ will be invalidated. Let $i$ be the instruction be $v = std::move(u)$. Then, \\[1ex]
\centerline{$ \varphi_i(d) = [d|v:d(u)|u:\{invalid\}]$}

    \item[Address-of operator] - when the address-of (\&)  operator is used, it creates a temporary variable that points to the operand of \&. Let $i$ be $\&v$. Then,\\[1ex]
\centerline{$
        \varphi_i(d) = 
        \begin{cases}
        [d|tmp:\{v\}] & \text{if} \; v \; \text{is a local variable} \\
        [d|tmp:\{global\}] & \text{if} \; v \; \text{is a global variable}
        \end{cases}
$}

    \item[\Pointer dereferencing] - when we dereference a \Pointer, a temporary variable may be created if the result is not a value. This is useful for cases when \Pointer to \Pointer types are used. Let $i$ be $*v$. Then,\\[1ex]
\centerline{$
        \varphi_i(d) = 
        \begin{cases}
        d & \text{if} \; *v \; \text{is a \Value} \\
        [d|tmp:d(d(v))\}] & \text{otherwise}
        \end{cases}
$}

    \item[Memory allocation] - when memory is allocated, a new \Owner is created to represent the memory zone returned by the allocate statement. Let $i$ be $\allocate(x)$. Then,\\[1ex]
\centerline{$ 
\varphi_i(d) = [d|\text{alloc}_x:\text{alloc}_x'] $}

    \item[Memory deallocation] - when memory is deallocated through a pointer $p$, all owners in its pset are invalidated. Let $i$ be $\deallocate(p)$. Then,\\[1ex]
\centerline{$\varphi_i(d) = \text{invalid}_{d(p)}(d)$}

    \item[Function calls] - when analyzing function calls, we assume that the annotated postconditions are true, and use them as the pset of the function output. Let $i = p=f()$ and the postcondition $post(f,S)$. Then,
    \[ \varphi_i(d) = [d|v:S] \]

    \item[Function definitions] - when analyzing function bodies, we assume that the annotated preconditions are true, and use them as the psets of the function parameters. Let $f(x)$ be a function annotated with $pre(x,S)$, Then,\\[1ex]
\centerline{$ \varphi_i(d) = [d|x:S] $}

    \item[All other instructions] - all other instructions that do not affect memory, there are no changes to the pmap. So for all other $i$,\\[1ex]
\centerline{$ \varphi_i(d) = d $}
\end{description}

After analyzing the whole CFG and calculating the fixpoint solution of the dataflow system, we have access to the $pmap$ of all nodes and can enforce the rules presented above in the following way:

\begin{itemize}
    \item It is an error to dereference an invalid pointer. If $i$ is $*p$ and $\invalid \in d(p)$, then we report the error.
    \item It is an error to dereference an \Owner that was moved from. If $i$ is  $*o$ and $\invalid \in d(o)$, then we report the error.
    \item It is an error to pass a \Pointer as a function argument if it is invalid, or violates the preconditions of the function. Let $f$ be a function annotated with $pre(p,S)$. At every function call $i = f(p)$, if $\invalid \in d(p)$, an error will be reported. In addition, if there is no $p' \in S$ such that $d(p) = d(p')$, then the precondition is violated and an error will be reported.
    \item It is an error to return a \Pointer from a function that is invalid or violates the postconditions of the function. Let $f$ be a function annotated with $post(f,S)$. At every statement that returns, $i = \text{return} \; p$, if $\invalid \in d(p)$, an error will be reported. In addition, if there is no $p' \in S$ such that $d(P) = d(p')$, then the postcondition is violated and an error will be reported
\end{itemize}

\begin{figure}
    \centering
    \includegraphics[width=0.7\linewidth]{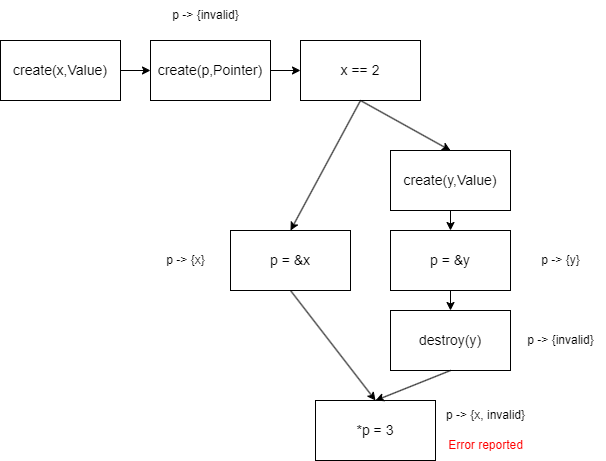}
    \caption{Example CFG analysis}
    \label{fig:cfgan}
\end{figure}

Figure \ref{fig:cfgan} shows an example analysis of the program in listing \ref{lst:transprog} annotated with the relevant changes in the $pmap$ entry of $p$.

\section{Implementation and Evaluation}

\subsection{Implementation}

The implementation of the framework consists of three main components: 
\begin{itemize}
    \item Parsing and processing a C++ source file received as input in order to transform it into the programming language used during the analysis.
    \item Creating the control flow graph and solving the dataflow system in order to calculate the $pmap$ at each node.
    \item Using the $pmap$s obtained after the previous step, identify and report any errors.
\end{itemize}

In addition to this, additional work has been done to improve error messages and help identify the source of errors faster. The main feature created with this purpose is the ability to track the moments when pointers are invalidated, and include these locations when an error is reported.

Because the main interest of our framework is not parsing the language, but performing static analysis on it, choosing an already implemented solution for this would be the best choice. Not only does this save resources because we do not need to implement a parser from scratch, but also it provides a well-tested, maintained, and well-documented parser which we can take advantage of.

The tool we chose for parsing the source files is Clang \cite{clang}. Clang is an open-source project that provides a language front-end and tooling infrastructure for languages in the C family. It is considered one of the three main C++ compilers and is widely used to build production-quality applications~\cite{cppstats}.

\subsection{Evaluation}

During each stage of development of the framework, several tests were run to evaluate the performance and accuracy of each component. Both handwritten, specific tests for certain functionalities, and real-world snippets and projects have been used to be able to accurately evaluate the framework.

\subsubsection*{Type classification tests}

Being one of the first steps of the analysis, type classification plays an important role and any misclassification will influence all further steps in the analysis, possibly leading to erroneous reporting of bugs. This is why we considered it necessary during development to keep a 100\% pass rate of tests in this category, as any errors here would make the evaluation of other features impossible. This also meant that the quality of the test had to be high, covering as many cases as possible

The test set consists of 30 tests and was created to cover every rule specified in Section~\ref{sec:types}, using multiple approaches. Both user-created and standard library types were used in the tests, in as many combinations as possible.

The framework passes all 30 tests, and the distribution of tests is presented in Table~\ref{tab:typeClass}.

\begin{table}[h!]
\centering
\caption{Type classification tests}
\label{tab:typeClass}
\begin{tabular}{|c|c|c|c|c|c|}
\hline
Type classification 1 & Total tests & User types & Std types & Both types & Pass rate \\
\hline
\Owner & 14 & 6 & 5 & 3 & 100\%\\
\hline
\Pointer & 10 & 3 & 5 & 2 & 100\%\\
\hline
\Value & 6 & 3 & 2 & 1 & 100\%\\
\hline
\end{tabular}
\end{table}

\subsubsection*{Error detection tests}

To evaluate the accuracy of the framework, we categorized the results of each error detection test as true positive, false positive, true negative, and false negative.

The most important metric to take into consideration is the number of false negatives. Our goal is to minimize this number, so that as few errors are missed as possible. The next priority is minimizing the number of false positives, as flooding the user with reports of inexistent errors might make them miss a true positive report.

The test set constructed for this evaluation aims to cover both erroneous as well as error-free code. The errors present should follow as many different patterns as possible. One important aspect is that different people write code in different manners, depending on their skills, experience, influences, and time invested. We included three different styles of writing code to try to emulate real-world codebases:
\begin{description}
    \item[Advanced] - this style of code is compliant with modern C++ best practices (see for example \cite{cppguidelines}), heavily uses the standard library, and designs clean and correct code. This would correspond to code written by a very experienced programmer who rarely makes mistakes.
    \item[Regular] - this style of code uses only the most well-known parts of the standard library while implementing other algorithms and data structures from scratch. This style would be the most common in real-world projects, corresponding to an experienced programmer, that knows and applies most best practices.
    \item[Basic] - this style of code barely uses the standard library, while possibly using C++ anti-patterns that may cause bugs in some situations. This style would be common among inexperienced programmers, or old codebases that have not been modernized.
\end{description}
The test suite consists of 102 tests and the results of the evaluation can be seen in Table~\ref{tab:errTest}.

\begin{table}[h!]
\centering
\caption{Error detection tests}
\label{tab:errTest}
\begin{tabular}{|c|c|c|c|c|}
\hline
Code style & Total tests & False Negative & False Positive & Accuracy \\
\hline
Advanced & 33 & 0 & 1 & 96.96\% \\
\hline
Regular & 35 & 3 & 2 & 85.7\%\\
\hline
Basic & 34 & 6 & 10 & 52.94\%\\
\hline
\end{tabular}
\end{table}

From the results, it can be concluded that the accuracy of error detection increases the more modern the code is. One cause of this is that by aligning with modern standards and writing code with best practices in mind, like const-correctness, using the standard library when possible, or designing classes with intuitive interfaces, the static analyzer can infer much more information about the context of the program. While the programmer can deduce most information from context, patterns or naming conventions, the compiler has no way doing this. Some examples of failed tests and potential solutions are presented below.

\subsubsection*{Analyzing Real-World Code}

In addition to evaluating the framework on small snippets of code, it is important to test its behavior in a real-world use case, on code bases of different sizes. We have chosen 5 projects of sizes from a few hundred lines to over 100000, and ran the analysis of them, checking the number of errors and validating whether they were correctly identified.

Project A is a small application written as a helper tool. It contains approximately 500 lines of modern C++ and implements simple functionalities like reading and writing files on the disk and manipulating strings.

Project B is a project that is currently in development and has around 4000 lines of code. It implements medium-complexity functionalities for filtering and grouping data from multiple sources.

Project C is a production application that is used as an interpreter for a domain-specific language. It is a medium-size code base of around 15 thousand lines of modern C++.

Project D is a complex application, containing over 40 thousand lines of code, written over a few years. Its features are of high complexity and include parsing data streams, extracting and correlating information, and using it to generate reports.

Project E is a very old code base, comprised of over 100 thousand lines of mostly C and C++98. It contains multiple modules, with very complex features and multiple interactions with operating system interfaces.

The results and the evaluation can be found in Table~\ref{tab:projErr} and Table~\ref{tab:projSpeed}.

\begin{table}[h!]
\centering
\caption{Error detection tests}
\label{tab:projErr}
\begin{tabular}{|c|c|c|c|c|}
\hline
Project & Coding style & Errors & False Positives & FP rate \\
\hline
A & Advanced & 0 & 0 & 0\% \\
\hline
B & Regular & 6 & 5 & 83.33\%\\
\hline
C & Advanced & 4 & 1 & 25\%\\
\hline
D & Regular & 105 & 105 & 100\%\\
\hline
E & Basic & 1038 & ?? & ??\%\\
\hline
\end{tabular}
\vspace{-2ex}
\end{table}

\begin{table}[h!]
\centering
\caption{Compilation speed tests}
\label{tab:projSpeed}
\begin{tabular}{|c|c|c|c|c|}
\hline
Project & Compilation & Compilation + Analysis & Delta & Slowdown \\
\hline
A & 53 ms & 68 ms & 13 ms & 24.5\% \\
\hline
B & 6272 ms & 9057 ms & 2785 ms & 44.4\%\\
\hline
C & 54 s & 76 s & 22 s & 40.7\%\\
\hline
D & 106 s & 139 s & 33 s & 31.1\%\\
\hline
E & 583 s & 721 s & 139 s & 23.8\%\\
\hline
\end{tabular}
\end{table}

It can be seen that the more modern and well-maintained the code is, the better the framework behaves. Although project C has almost 4 times as many lines of code as project B, the false positive rate is much lower. This may be attributed to the more modern design of the code and the much heavier use of standard library functionalities that enable the analysis to accurately infer the needed information. 

The results of project D also bring up a potentially important issue: all 105 reported errors were false positives, having the same root cause: not annotating functions and variables as const when needed. 

Project E generated so many errors caused by a variety of anti-patterns, legacy code, and lack of refactorization when needed, that it was impossible to manually validate each one. This may raise a problem when the framework has to analyze huge, legacy code bases: flooding the user with errors, most of which are false positives, caused mostly by legacy code, which would require a significant time investment to solve.

\subsubsection*{Discussion}

The evaluation of the framework showed good results, but at the same time, the false positives and false negatives show that there are more improvements to be made. The main causes of test failures we have been able to identify are:
\begin{itemize}
    \item User-defined types that behave like an Owner, but don't fully follow the specification in our framework.
    \item Types that have shared-ownership behaviour (for example $std::shared_ptr$.
    \item Functions that have no side-effects, but are not annotated $const$.
\end{itemize}

Further improvements of type classification rules and program analysis are needed to reduce these failures.

\section{Conclusions and Future Work}

\paragraph*{Conclusions}

The goal when the development of the framework started was to be able to prevent some exploits from ever becoming available to the public by detecting the errors that would cause the vulnerability during the development process, and warning the programmer so that it can be fixed.

The framework we developed is able to take any C++ source file as input, using even the most recently standardized features, facilitated by Clang's parser and AST. Then, it will transform the AST into a control flow graph, adapting the original program to the language used during the analysis, by transforming its types, instructions and functions. After the creation of the CFG, a dataflow system is defined and solved for each function in the program, to extract the necessary information about the points-to maps at each node. Finally, using this information, a set of rules is enforced, and errors are reported to the programmer through intuitive messages, that contain the location of the bug, in addition to potential previous instructions that may cause it.

Multiple tests were performed in order to evaluate the accuracy and speed of the framework. While most tests were successful, some of them showed some potential problems when analyzing big codebases. The main one is that for human programmers, it may be easy to understand information from the context of the code, without the need of explicit annotations, while for static analyzers that is impossible.

It can be concluded that our framework achieved its goals, as it is able to detect multiple patterns of use-after-free bugs, and correctly report the errors to the programmer.

\paragraph*{Future Work}

While we can say that the framework was able to achieve its goals, there are many improvements that can be made. Some of the most important are:

\begin{itemize}
    \item Modifying the type classification rules to be able to recognise shared ownership semantics. Some cases of false positives are caused by types that have the behavior of a shared \Owner, which invalidate all Pointers when a single \Owner is destroyed, while in reality the Pointers remain valid.
    \item Improve type classification to be able to identify types which represent Owners, but don't fall into any of the categories mentioned in our rules.
    \item Implement mechanisms that are able to deduce some information without it being explicitly mentioned. One good example is being able to identify whether a variable or function can be considered const.
    \item Research interprocedural analysis to be able to detect much more patterns of errors.
\end{itemize}

\nocite{*}
\bibliographystyle{eptcs}
\bibliography{generic}

\begin{thebibliography}{10}
\providecommand{\bibitemdeclare}[2]{}
\providecommand{\surnamestart}{}
\providecommand{\surnameend}{}
\providecommand{\urlprefix}{Available at }
\providecommand{\url}[1]{\texttt{#1}}
\providecommand{\href}[2]{\texttt{#2}}
\providecommand{\urlalt}[2]{\href{#1}{#2}}
\providecommand{\doi}[1]{doi:\urlalt{https://doi.org/#1}{#1}}
\providecommand{\eprint}[1]{arXiv:\urlalt{https://arxiv.org/abs/#1}{#1}}
\providecommand{\bibinfo}[2]{#2}

\bibitemdeclare{misc}{cppguidelines}
\bibitem{cppguidelines}
\emph{\bibinfo{title}{{C++ Core Guidelines}}}.
\newblock
  \urlprefix\url{https://isocpp.github.io/CppCoreGuidelines/CppCoreGuidelines}.
\newblock
  \bibinfo{note}{\url{https://isocpp.github.io/CppCoreGuidelines/CppCoreGuidelines}
  last visited in June 2024}.

\bibitemdeclare{misc}{clang}
\bibitem{clang}
\emph{\bibinfo{title}{{Clang ecosystem}}}.
\newblock \urlprefix\url{https://clang.llvm.org/}.
\newblock \bibinfo{note}{\url{https://clang.llvm.org/} last visited in June
  2024}.

\bibitemdeclare{misc}{clangcheck}
\bibitem{clangcheck}
\emph{\bibinfo{title}{{Clang static analyzer checkers}}}.
\newblock \urlprefix\url{https://clang.llvm.org/docs/analyzer/checkers.html}.
\newblock
  \bibinfo{note}{\url{https://clang.llvm.org/docs/analyzer/checkers.html} last
  visited in June 2024}.

\bibitemdeclare{misc}{clion}
\bibitem{clion}
\emph{\bibinfo{title}{{CLion warnings}}}.
\newblock
  \urlprefix\url{https://www.jetbrains.com/help/clion/list-of-c-cpp-inspections.html}.
\newblock
  \bibinfo{note}{\url{https://www.jetbrains.com/help/clion/list-of-c-cpp-inspections.html}
  last visited in June 2024}.

\bibitemdeclare{misc}{bluekeep}
\bibitem{bluekeep}
\emph{\bibinfo{title}{{The Bluekeep bug}}}.
\newblock \urlprefix\url{https://en.wikipedia.org/wiki/BlueKeep}.
\newblock \bibinfo{note}{\url{https://en.wikipedia.org/wiki/BlueKeep} last
  visited in June 2024}.

\bibitemdeclare{misc}{eternalblue}
\bibitem{eternalblue}
\emph{\bibinfo{title}{{The EternalBlue bug}}}.
\newblock \urlprefix\url{https://en.wikipedia.org/wiki/EternalBlue}.
\newblock \bibinfo{note}{\url{https://en.wikipedia.org/wiki/EternalBlue} last
  visited in June 2024}.

\bibitemdeclare{misc}{heartbleed}
\bibitem{heartbleed}
\emph{\bibinfo{title}{{The Heartbleed bug}}}.
\newblock \urlprefix\url{https://heartbleed.com/}.
\newblock \bibinfo{note}{\url{https://heartbleed.com/} last visited in June
  2024}.

\bibitemdeclare{misc}{cppstats}
\bibitem{cppstats}
\emph{\bibinfo{title}{{The State of Developer Ecosystem 2023 - Jetbrains}}}.
\newblock \urlprefix\url{https://www.jetbrains.com/lp/devecosystem-2023/cpp/}.
\newblock
  \bibinfo{note}{\url{https://www.jetbrains.com/lp/devecosystem-2023/cpp/} last
  visited in June 2024}.

\bibitemdeclare{inproceedings}{afl}
\bibitem{afl}
\bibinfo{author}{Andrea \surnamestart Fioraldi\surnameend},
  \bibinfo{author}{Dominik~Christian \surnamestart Maier\surnameend},
  \bibinfo{author}{Heiko \surnamestart Ei{\ss}feldt\surnameend} \&
  \bibinfo{author}{Marc \surnamestart Heuse\surnameend} (\bibinfo{year}{2020}):
  \emph{\bibinfo{title}{{AFL++} : Combining Incremental Steps of Fuzzing
  Research}}.
\newblock In \bibinfo{editor}{Yuval \surnamestart Yarom\surnameend} \&
  \bibinfo{editor}{Sarah \surnamestart Zennou\surnameend}, editors: {\slshape
  \bibinfo{booktitle}{14th {USENIX} Workshop on Offensive Technologies, {WOOT}
  2020, August 11, 2020}}, \bibinfo{publisher}{{USENIX} Association},
  \doi{10.5555/3488877.3488887}.
\newblock
  \urlprefix\url{https://www.usenix.org/conference/woot20/presentation/fioraldi}.

\bibitemdeclare{book}{compDesign}
\bibitem{compDesign}
\bibinfo{author}{Dick \surnamestart Grune\surnameend},
  \bibinfo{author}{Henri~E. \surnamestart Bal\surnameend},
  \bibinfo{author}{Ceriel J.~H. \surnamestart Jacobs\surnameend} \&
  \bibinfo{author}{Koen \surnamestart Langendoen\surnameend}
  (\bibinfo{year}{2002}): \emph{\bibinfo{title}{Modern Compiler Design}}.
\newblock \bibinfo{publisher}{John Wiley}.

\bibitemdeclare{misc}{lifetime}
\bibitem{lifetime}
\bibinfo{author}{\surnamestart {Herb Sutter}\surnameend}:
  \emph{\bibinfo{title}{{Lifetime safety: Preventing common dangling}}}.
\newblock
  \urlprefix\url{https://github.com/isocpp/CppCoreGuidelines/blob/master/docs/Lifetime.pdf}.
\newblock
  \bibinfo{note}{\url{https://github.com/isocpp/CppCoreGuidelines/blob/master/docs/Lifetime.pdf}
  last visited in June 2024}.

\bibitemdeclare{book}{cppstd}
\bibitem{cppstd}
\bibinfo{author}{\surnamestart {International Organization for
  Standardization}\surnameend} (\bibinfo{year}{2020}):
  \emph{\bibinfo{title}{Programming Languages - {C++}}}, \bibinfo{edition}{5th}
  edition.
\newblock \bibinfo{series}{ISO/IEC 14882:2020},
  \bibinfo{publisher}{International Organization for Standardization},
  \bibinfo{address}{Geneva, Switzerland}.

\bibitemdeclare{inbook}{notpetya}
\bibitem{notpetya}
\bibinfo{author}{Csaba \surnamestart Krasznay\surnameend}
  (\bibinfo{year}{2020}): \emph{\bibinfo{title}{Case Study: The NotPetya
  Campaign}}, pp. \bibinfo{pages}{485--499}.

\bibitemdeclare{inproceedings}{wannacry}
\bibitem{wannacry}
\bibinfo{author}{M.~Satheesh \surnamestart Kumar\surnameend},
  \bibinfo{author}{Jalel \surnamestart Ben{-}Othman\surnameend} \&
  \bibinfo{author}{K.~G. \surnamestart Srinivasagan\surnameend}
  (\bibinfo{year}{2018}): \emph{\bibinfo{title}{An Investigation on Wannacry
  Ransomware and its Detection}}.
\newblock In: {\slshape \bibinfo{booktitle}{2018 {IEEE} Symposium on Computers
  and Communications, {ISCC} 2018, Natal, Brazil, June 25-28, 2018}},
  \bibinfo{publisher}{{IEEE}}, pp. \bibinfo{pages}{1--6},
  \doi{10.1109/ISCC.2018.8538354}.

\bibitemdeclare{book}{flexbison}
\bibitem{flexbison}
\bibinfo{author}{John~R. \surnamestart Levine\surnameend}
  (\bibinfo{year}{2009}): \emph{\bibinfo{title}{flex and bison - Unix text
  processing tools}}.
\newblock \bibinfo{publisher}{O'Reilly}.
\newblock
  \urlprefix\url{http://www.oreilly.de/catalog/9780596155971/index.html}.

\bibitemdeclare{inproceedings}{valgrind}
\bibitem{valgrind}
\bibinfo{author}{Nicholas \surnamestart Nethercote\surnameend} \&
  \bibinfo{author}{Julian \surnamestart Seward\surnameend}
  (\bibinfo{year}{2007}): \emph{\bibinfo{title}{Valgrind: a framework for
  heavyweight dynamic binary instrumentation}}.
\newblock In \bibinfo{editor}{Jeanne \surnamestart Ferrante\surnameend} \&
  \bibinfo{editor}{Kathryn~S. \surnamestart McKinley\surnameend}, editors:
  {\slshape \bibinfo{booktitle}{Proceedings of the {ACM} {SIGPLAN} 2007
  Conference on Programming Language Design and Implementation, San Diego,
  California, USA, June 10-13, 2007}}, \bibinfo{publisher}{{ACM}}, pp.
  \bibinfo{pages}{89--100}, \doi{10.1145/1250734.1250746}.

\bibitemdeclare{book}{progAnalysis}
\bibitem{progAnalysis}
\bibinfo{author}{Flemming \surnamestart Nielson\surnameend},
  \bibinfo{author}{Hanne~Riis \surnamestart Nielson\surnameend} \&
  \bibinfo{author}{Chris \surnamestart Hankin\surnameend}
  (\bibinfo{year}{1999}): \emph{\bibinfo{title}{Principles of program
  analysis}}.
\newblock \bibinfo{publisher}{Springer}, \doi{10.1007/978-3-662-03811-6}.

\bibitemdeclare{inproceedings}{asan}
\bibitem{asan}
\bibinfo{author}{Konstantin \surnamestart Serebryany\surnameend},
  \bibinfo{author}{Derek \surnamestart Bruening\surnameend},
  \bibinfo{author}{Alexander \surnamestart Potapenko\surnameend} \&
  \bibinfo{author}{Dmitriy \surnamestart Vyukov\surnameend}
  (\bibinfo{year}{2012}): \emph{\bibinfo{title}{AddressSanitizer: {A} Fast
  Address Sanity Checker}}.
\newblock In \bibinfo{editor}{Gernot \surnamestart Heiser\surnameend} \&
  \bibinfo{editor}{Wilson~C. \surnamestart Hsieh\surnameend}, editors:
  {\slshape \bibinfo{booktitle}{2012 {USENIX} Annual Technical Conference,
  Boston, MA, USA, June 13-15, 2012}}, \bibinfo{publisher}{{USENIX}
  Association}, pp. \bibinfo{pages}{309--318}, \doi{10.5555/2342821.2342849}.
\newblock
  \urlprefix\url{https://www.usenix.org/conference/atc12/technical-sessions/presentation/serebryany}.

\bibitemdeclare{inproceedings}{fuzz}
\bibitem{fuzz}
\bibinfo{author}{Kosta \surnamestart Serebryany\surnameend}
  (\bibinfo{year}{2016}): \emph{\bibinfo{title}{Continuous Fuzzing with
  libFuzzer and AddressSanitizer}}.
\newblock In: {\slshape \bibinfo{booktitle}{2016 IEEE Cybersecurity Development
  (SecDev)}}, pp. \bibinfo{pages}{157--157}, \doi{10.1109/SecDev.2016.043}.

\bibitemdeclare{inproceedings}{catastrofa}
\bibitem{catastrofa}
\bibinfo{author}{W.~Eric \surnamestart Wong\surnameend},
  \bibinfo{author}{Vidroha \surnamestart Debroy\surnameend},
  \bibinfo{author}{Adithya \surnamestart Surampudi\surnameend},
  \bibinfo{author}{HyeonJeong \surnamestart Kim\surnameend} \&
  \bibinfo{author}{Michael~F. \surnamestart Siok\surnameend}
  (\bibinfo{year}{2010}): \emph{\bibinfo{title}{Recent Catastrophic Accidents:
  Investigating How Software was Responsible}}.
\newblock In: {\slshape \bibinfo{booktitle}{Fourth International Conference on
  Secure Software Integration and Reliability Improvement, {SSIRI} 2010,
  Singapore, June 9-11, 2010}}, \bibinfo{publisher}{{IEEE} Computer Society},
  pp. \bibinfo{pages}{14--22}, \doi{10.1109/SSIRI.2010.38}.

\end{thebibliography}
\end{document}